\documentclass[twocolumn]{revtex4}
\usepackage{graphicx}
\usepackage{dcolumn}
\usepackage{bm}
\usepackage{multirow}
\usepackage[english]{babel}
\usepackage{wrapfig}
\usepackage{epsfig}
\usepackage{color}
\usepackage{amsmath}

\def\B{\mathop{\mathcal B}}

\def\FIG#1#2#3#4{%
  \begin{figure}
    \begin{center}%
      \includegraphics*[#4]{#2}%
      \caption{\label{#1}#3}%
  \end{center}
\end{figure}%
}

\begin{document}

\title{Flame fronts in Supernovae Ia and their pulsational stability}

\author{S.I. Glazyrin}
\email{glazyrin@itep.ru}
\affiliation{Institute for Theoretical and Experimental Physics, Moscow 117218, Russia}
\author{S.I. Blinnikov}
\email{sergei.blinnikov@itep.ru}
\affiliation{Institute for Theoretical and Experimental Physics,
  Moscow 117218, Russia}
\affiliation{Novosibirsk State University, Novosibirsk 630090, Russia}
\affiliation{VNIIA, Moscow 127055, Russia}
\author{A.D. Dolgov}
\email{dolgov@fe.infn.it}
\affiliation{Institute for Theoretical and Experimental Physics,
  Moscow 117218, Russia}
\affiliation{Novosibirsk State University, Novosibirsk 630090, Russia}
\affiliation{University of Ferrara and INFN, Ferrara 44100, Italy}

\begin{abstract}
  The structure of the deflagration burning front in type Ia
  supernovae is  considered. The parameters of the flame are obtained: its
  normal velocity and thickness. The results are in good agreement
  with the previous works of different authors. The problem of pulsation instability of
  the flame, subject to plane perturbations, is studied.
  First, with the artificial system with switched--off hydrodynamics
  the possibility of secondary reactions to stabilize the front is shown.
  Second, with account of hydrodynamics, realistic EOS and thermal
  conduction we can obtain pulsations when Zeldovich number was
  artificially increased. The critical Zeldovich numbers are
  presented. These results show the stability of the flame  
  in type Ia supernovae against pulsations because its effective
  Zeldovich number is small.
\end{abstract}

\maketitle

\section{Introduction}

Supernovae explosions are among  the most spectacular  events in the
Universe: their energy release significantly mixes the interstellar medium and
acts like a driving force in gas dynamics of galaxies and production of cosmic rays.
The luminosity of an exploding star becomes 
comparable with the luminosity of the progenitor galaxy, and allows 
to observe processes in the most distant regions of the Universe. 
If regular features of the supernova 
explosions are found for any subtype of supernovae,
it could open a new way to measure cosmological distances and the values of the
cosmological parameters.

Despite a long history of investigations of these events the complete
understanding of underlying physics is still missing.
There are several kind of supernova explosions of different types of the 
progenitor stars with absolutely different physical phenomena behind them.
Here we will consider only one subtype of explosions, namely the
thermonuclear explosions. These kind of supernovae is called 
the  supernovae of type Ia, SNIa.
Analysis of the observational data indicates that the explosion is induced  by the
thermonuclear burning of premixed carbon--oxygen fuel. Such phenomenon 
usually takes place in degenerate stars, white dwarfs.

The mode of the explosive nuclear burning in supernovae
is still a controversial issue, in spite of many years
of the research in the field. Four decades ago,
\citet{Arn} was the first to consider supersonic
combustion, i.e. detonation, in supernovae. Later,
 \citet{IIC} obtained a sub-sonic flame (deflagration)
propagating in spontaneous regime with pulsations and a
subsequent transition to detonation, while \citet{NSN}
considered the deflagration propagating due to convective heat transfer.
Both detonation and deflagration have their merits and problems
in explaining the supernova phenomenon (see, e.g. \cite{WW}). It
is not clear if detonation succeeds or fails to develop, but it is
clear that in any case the combustion
must be much faster than it is suggested by the analysis of the propagation
of a laminar one-dimensional flame.

From microscopic point of view one-dimensional nuclear
flame is a wave described  essentially in the same way
as it was done by
\citet{ZFK}
in spite of
complications introduced by nuclear kinetics and very high
conductivity of dense presupernova matter.
It is found that the conductive flame propagates in  presupernova with the
speed which is too slow
to explain the supernova outburst correctly since
the flame Mach number is of the order of one percent or less
\citep{TimmesWoosley_ApJ_1992}.

The fuel consumption can naturally be accelerated by the development
of the instabilities inherent to the flame front.
As it is explained in the classical paper by 
\citet{Landau,Landau2}, the hydrodynamic instability leads to wrinkling
or roughening of the front surface, and hence to an increase of its area with
respect to the smooth front and consequently to an acceleration of
the flame propagation.
In some cases, observed in laboratory experiments like \cite{Gostintsev}, when the LD
instability is really strong (large density jump accross the flame
front, and the flame is freely expanding) such instabilities can lead
to a transition from the regime of slow
flame propagation  to the  regime of detonation.
Since the flame propagates in gravitational field, and the burned ashes
have lower density than the unburned fuel, the Rayleigh--Taylor (RT)
instability is often considered to be the dominant instability governing
the corrugation of the front \cite{MA,wooa,woob,LivA,kho}.
The RT instability creates turbulent cascade providing  an
acceleration of the flame front.
However it leads to additional difficulties in modeling the SN
event \cite{NHt,Niem,Nwoo,WoosleyKersteinEtAl_ApJ_2010,AspdenEtAl_ApJ_2008}.

It is well known \citep{Landau,Landau2,LL,Will} 
that large portions of a slow
planar flame front are unstable with respect to the large scale bending.
This universal instability is called the Landau-Darrieus (LD)
instability. 
For the wavelengths much longer than the flame thickness
it does not depend,
on complex processes which take place in the burning zone.
Development of the LD instability depends only on the
sign of $\Delta\rho = \rho_{\rm u} -\rho_{\rm b}$,
where $\rho_{\rm u}$ and
$\rho_{\rm b}$ are the densities of the unburned and burned ``gases''
respectively. The LD instability of the planar flame fronts with respect to
large scale bending takes place if and only if $\Delta\rho >0$.

The LD instability plays an important role in many physical phenomena
such as the usual chemical burning of gases,
explosive boiling of liquids \citep{Frost},
electroweak phase transitions \citep{KamFr}, dynamics of thermally bistable
gas \citep{AMS2}, and thermonuclear burning in
supernovae  \citep{BSW,NHld}. The detailed consideration the of non-linear
stage of the LD instability and the calculation of the fractal dimension
of the flame front for this case
is given by 
\citet{BS,Joulin_PRE_1994}. It is
interesting, and somewhat puzzling, that a similar
dependence of the flame fractal dimension on the density discontinuity was found
in the 3D SPH simulations of the flame subject to the RT instability
\citep{BrGar}.

Both LD and RT instabilities develop on scales much larger than
the flame thickness and they can be successfully studied in the
approximation of the discontinuous front. This approximation is
not valid for another instability, first discovered by \citet{Zelpw} 
in his investigation of the powder combustion.
This instability originated from
a strong temperature dependence of the reactions rates. As a result of  that local
fluctuations of the heating rate caused by the temperature fluctuations
cannot be controlled by  thermal conduction.
This phenomenon can lead to a pulsating regime of the front
propagation and to a renormalization of the mean front velocity \cite{W}.
Such instability can develop
even for one-dimensional perturbations when the plane front preserves
its shape. We denote it as TP (thermal-pulsational) instability.
After publication of paper by \citet{BarZI} TP instability
was studied quantitatively in many works
(for the list of references see the book \cite{Zel}).

A very nice review on the SNIa physics is given in \cite{HillebrandtNiemeyer_astro_ph_0006305}  
(see also \cite{RopkeHillebrandtBlinnikov2006,RopkeHillebrandtEtAl_ApJ_2007}).
There exist several scenarios of SNIa explosions. The most popular are the following:
the single--degenerate scenario, the double--degenerate scenario, and the
sub-Chandrasekhar mass explosion. In this paper the single degenerate scenario is
considered.

The paper is organized as follows. In Section \ref{sec:model}  
a model of a white dwarf explosion is presented and the physical conditions are
discussed. In  Section
\ref{sec:flame_prop_1d} the stationary propagating flame is considered analytically and
numerically. The dependence of the results on nuclear reaction
network is discussed. In Section \ref{sec:th_dyn_ins} artificial
systems with switched off hydrodynamics are considered. The
effect of secondary reaction on pusations is considered.
In Section \ref{sec:front_puls} the stability of the
flame under conditions close to those in a white dwarf is
considered. We show that pulsations could exist in this system when
the Zeldovich number is aftificially increased. After it we make
conclusions about stability of real flames.


\section{The model}
\label{sec:model}

According to the single degenerate scenario a binary stellar system,
which is a progenitor of the supernova, consists
of a white dwarf (WD) and a non degenerate star. During accretion of matter
on the WD it approaches the Chandrasekhar mass limit and at some moment becomes
unstable. In the language of equation of state it happens because the
adiabatic exponent approaches the critical value
$\gamma\approx 4/3$. In the course of  this process the temperature in the centre of
the WD rises and nuclear burning of matter is ignited. According to evolutionary models, 
the matter consists of  degenerate $^{12}$C and $^{16}$O nuclei.
The temperature of ignition depends on the matter
density and can be found in \citep{Potekhin2012}. For $\rho\sim 10^9$
g/cm$^3$ it is about $T\sim 10^8$ K. But this burning is very slow and does not
propagate outwards from the centre until  its energy release stops to be 
compensated for by various losses (the most significant are the neutrino
losses). The dynamical stage sets in later,
when $T$ rises up to $10^9$ K and the flame is born.
This is the beginning of the supernova explosion.
The flame starts to propagate from the centre of the star to its
surface. The regime of the flame propagation is under intensive investigation 
but the answer is not yet found.
Still it is not fully unknown, the observations imply some limitations on it.
There exist two types of stationary regimes:
1) deflagration, when the flame propagation 
velocity is small compared to the speed of sound and the flame is driven
by dissipative effects: thermo-conductivity or diffusion; 
2)detonation, when a supersonic wave propagates with the shock front where the temperature jumps up
drastically  leading to fast burning. 
If the burning of the whole star
proceeds in the detonation regime, it burns up to Fe-peak elements. However, it
contradicts observations: in a real supernova about half of the star should consist of
the intermediate elements.
Pure deflagration regime does not succeed too: the star expands with velocity, 
which is faster than the flame velocity so the temperature drastically drops down
 and all the burning terminates. 
The only feasible successful regime is the mixed one when the flame starts with 
deflagration in high density matter, where the expansion coefficient
is small. Then it somehow accelerates at the radius, which is usually characterized
by some critical density, and passes to detonation. This model
successfully  explains all the parameters of the explosion at the expence of
one tuning parameter, critical density,  $\rho_{\rm crit}$. From comparison of 
simulations with observations it is found
that $\rho_{\rm crit}\sim 2\cdot 10^7$ g/cm$^3$ 
(see, e.g. \cite{KhokhlovOranWheeler_ApJ_1997,HillebrandtNiemeyer_astro_ph_0006305}).
But the global problem in SNIa physics (that is surely not solved in this paper)
is to construct the model of the explosion with no tuning
parameters. One should calculate the critical density of the burning regime
transition from the first principles. This is the problem for the future.


Let us discuss physical conditions in a white dwarf assuming that its
mass is close to the Chandrasekhar mass limit. The density in the centre of WD
is $\rho\sim 10^9$ g/cm$^3$. The chemical composition is mainly
$^{12}$C and $^{16}$O. Hence  it can be shown that the temperature at which the flame starts
burning is about $T\sim 10^9$ K.
Two main physical processes that matter in the SNIa explosions are
thermo-conductivity and nuclear reactions (burning).
In this case ions are non-degenerate and non-relativistic, electrons on the contrary are
strongly or semi-degenerate and relativistic.
EOS for such matter is presented in \cite{NadyozhinEtAl_ApJS_1996} and
takes into account ions and all degrees of degeneracy of electrons and
photons.

Under these conditions the thermo-conductivity plays the leading role 
with respect to  other dissipative processes. There are two components of the
conductivity, radiative and electronic ones. The electronic thermo-conductivity
was calculated in \cite{YakovlevUrpin_SovietAstronomy_1980, GlazyrinBlinnikov_JPhysA_2010}. 
 An approximate equations for
radiative thermo-conductivity are presented in \cite{Iben_ApJ_1975}.
We evaluate the magnitude of the diffusion effects in the star under scrutiny as follows,
Despite the fact that in the white dwarf the matter is well
mixed, we estimate the value of the ion diffusion coefficient as:
\begin{equation}
  D\sim \lambda_i v_T\,,
\end{equation}
where $v_T$ is the thermal speed of the ions and $\lambda_i$ is the ion mean
free-path, which may be crudely estimated as the inter-ion distance.
So the Lewis number, the relation between thermoconductivity (the
coefficient is $\kappa$) and
diffusion, is about ${\rm Le}\equiv \kappa/\rho C_p D=10^4$. Here
$C_p$ is heat capacity under constant pressure. This
result can be
easily explained by the differences of velocities of the relativistic
particles, which contribute to
thermoconductivity, and the non-relativistic ones, which contribute to diffusion.
The shear viscosity can be written as:
\begin{equation}
  \eta=mn\lambda_i v_T.
\end{equation}
Where $m$ and $n$ are mass and concentration of ions.
So the Prandl number, the relation between thermoconductivity and
viscosity, is ${\rm Pr}\equiv C_p\eta/\kappa=10^{-4}$.

At the temperatures specified above, the nuclear reactions proceed in a branched network
with a lot of isotopes. Let us assume for simplicity that WD
consists only of $^{12}$C. Then the first reaction in the nuclear
network is $^{12}$C+$^{12}$C$\rightarrow^{24}$Mg$^*$, where  Mg$^*$ is 
an exited nuclei, which is unstable and decays into 3 channels with $p$, $n$, or
$\alpha$ in the final states. Its caloricity is $q=5.6\cdot 10^{17}$~erg/g, so it
leads to a significant temperature rise, up to $T\sim 10^{10}$ K, and
can provoke further burning. In this paper two variants of nuclear
network are considered:
\begin{enumerate}
\item A simplified network with only one reaction
    $^{12}$C+$^{12}$C$\rightarrow^{24}$Mg$^{*}$ (its rate can be found
    in \cite{CaughlanFowler_AtomicDataAndNuclDataTab_1988}). According
    to the fact that it is the first reaction in the network and due to the
    electromagnetic nature of this reaction it could be a good
    approximation for the whole network, so all
    kinetics would be determined by this reaction. 
    The complete burning up to Ni could also be modeled in this
    framework by fixing the
    rate and changing the caloricity to $q=9.2\cdot
    10^{17}$~erg/g. The explicit expressions for species production
    rates and energy generation for this network are:
    \begin{eqnarray}
      R_{\rm C12}&=&-F_{\rm scr}A_{\rm C12}^{-1}\rho X_{\rm C12}^2R(T),\\
      R_{\rm Mg24}&=&-R_{\rm C12},\nonumber\\
      \dot{S}&=&qR_{\rm Mg24},\nonumber\\
      R(T)&=&4.27\cdot
      10^{26}\frac{T_{9A}^{5/6}}{T_9^{1.5}}e^{-84.165/T_{9A}^{1/3}-2.12\cdot
      10^{-3}T_9^3},\nonumber\\
      T_{9A}&\equiv&\frac{T_9}{1+0.0396T_9}.\nonumber
    \end{eqnarray}
    For definitions see system (\ref{sys:hydro}). $A_{\rm C12}$ is
    the atomic mass of $^{12}$C, $F_{\rm scr}$ is the screening factor
    (see below), $T_9$ is temperature in units of $10^9$ K.
\item $\alpha$-chain with 13 isotopes from \cite{aprox13}. We  call
  it ``{\sc aprox13}''. This nuclear network is supplied with the code
  that calculates $R_i$ and $\dot{S}$.
\end{enumerate}
We make an additional simulations with initial $^{16}$O
composition using {\sc aprox13}.
Because of the high matter density the degeneracy parameter 
is not small, $\Gamma\sim E_{\rm coul}/E_{\rm kin}\sim 1$ (where
$E_{\rm coul}$ is a typical Coulomb energy per ion, and $E_{\rm kin}$
is a typical kinetic energy per ion),
so the electron nuclear screening effects
should be taken into account. The accurate screening
factor $F_{\rm scr}$ in a non-ideal gas can be obtained only by Monte--Carlo
simulations, and is presented, e.g.,  in \citep{ChabrierPotekhin_PRE_1998}. We should emphasize
that many-orders of magnitude discrepancy in the screening factor can arise because of incorrect
definitions, see details in \citep{GlazyrinBlinnikov_JPhysA_2010}. In
our simulations the screening factor ranges from $F_{\rm scr}\approx
10$ in unburned matter to $F_{\rm scr}\approx 1$ in the ashes
($F_{\rm scr}$ exponentially depends on $1/T$).

Based on the consideration presented above we conclude that the stellar explosion can be described by the
following system of hydrodynamic equations:
\begin{eqnarray}
  &&\frac{d\rho}{dt}=-\rho\frac{\partial v}{\partial x}, \nonumber\\
  &&\frac{dX_i}{dt}=R_{i},\nonumber\\
  &&\frac{dv}{dt}=-\frac{1}{\rho}\frac{\partial p}{\partial x},\label{sys:hydro}\\
  &&\frac{d\epsilon}{dt}=-\frac{p}{\rho}\frac{\partial v}{\partial x}-\frac{1}{\rho}\frac{\partial Q}{\partial x}+\dot{S},\nonumber\\
  &&Q=-\kappa\partial_x T,\nonumber\\
  &&\dot{S}=\sum\limits R_iB_i,\nonumber\\
  &&p=p(\rho, X_i, \epsilon),\nonumber
\end{eqnarray}
which we need to solve.
Here $X_i\equiv\rho_i/\rho$ are nuclide mass fractions ($i$=C12, Mg24 etc.), $\epsilon$ is internal energy per unit mass, $R_i$ is the
rate of production of species $i$, $\dot{S}$ -- energy generation
rate, $B_i$ -- binding energies of species $i$.

Though we consider processes in a star, the system whose dynamics is
crucially affected by gravity, the gravitational force is not included
in the system (\ref{sys:hydro}). As it will be seen further, the
thickness of the flame is negligible in comparison with scales of
variation of gravitational force. So gravity does not have an impact on flame
structure and all processes on the spatial scale of flame thickness.

\section{Properties of one-dimensional flame}
\label{sec:flame_prop_1d}
\subsection{Formulation of the Problem}

The main questions we pose here are the following: what is the normal velocity of flame
propagation, what is the expansion coefficient of matter over the flame front, and what
are the physical parameters of matter after the flame traversal.
To answer these questions we should study microphysics of the plane burning front
 taking into account the processes with characteristic scale of the order of the 
  front thickness.
  
Let us make some analytical estimations. For the deflagration regime
the characteristic times of burning $\tau_{\rm nucl}$ and heat
transfer $\tau_{\rm cond}$ over the front are respectively:
\begin{equation}
  \tau_{\rm nucl}=\frac{\rho q}{\dot{S}},~~\tau_{\rm
    cond}=\frac{d^2}{\kappa}.
  \label{eq:t_nucl_cond}
\end{equation}
If the flame is stationary, these times should be equal, and we can estimate the flame
thickness $d$ and velocity as:
\begin{equation}
  d=\sqrt{\kappa\frac{\rho q}{\dot{S}}},~~v=\frac{d}{\tau_{\rm
      cond}}=\sqrt{\kappa\frac{\dot{S}}{\rho q}}\label{eq:theor_vel}.
\end{equation}
In table \ref{tab:theor_vel} the estimated  velocity values are presented for
different conditions in WD, when the rate of burning $\dot{S}$ is
determined by the single reaction proposed earlier, (2$^{12}$C$\rightarrow^{24}$Mg$^*$).

\begin{table}
\caption{\label{tab:theor_vel}Theoretical estimate of the front velocity
  and thickness with 2$^{12}$C$\rightarrow^{24}$Mg$^*$ (Eqns. (\ref{eq:t_nucl_cond})--(\ref{eq:theor_vel})):}
\begin{center}
\begin{tabular}{|c|c|c|}
  \hline
  $\rho$, g/cm$^3$ & $v$, km/s & $\Delta x$, cm\\
  \hline
  $2\cdot 10^8$ & 203 & $2.5\cdot 10^{-5}$ \\
  $7\cdot 10^8$ & 742 & $3.0\cdot 10^{-6}$ \\
  $2\cdot 10^9$ & 1300 & $1.0\cdot 10^{-6}$ \\
  \hline
\end{tabular}
\end{center}
\end{table}

The exact front parameters can be obtained only by the numerical
simulations of system (\ref{sys:hydro}) in 1D.
For general rates, $R_{ij}$, of the reactions in system~(\ref{sys:hydro}) the density and temperature
distributions in a steady burning wave are unknown, so the initial conditions with the steady flame
could not be imposed. We simulate the ignition of the 
the flame by a warm wall and try to
eliminate all interfering perturbations. We expect that the
stationary flame would naturally appear in such conditions.

The numerical calculations are made for the following model conditions. The region of interest
is supposed to consist of uniformly distributed matter. 
For simplicity it is assumed to be be $^{12}$C with
density $\rho_0$ and  temperature $T_0$, chosen so that the characteristic time of
the flame burning is much larger than the dynamical
time in the problem under consideration. In this case the ignition is controlled by 
the boundaries. The right hand side boundary condition sets a constant external
pressure. The left hand side boundary condition is a hard wall with the vanishing velocity, $v=0$, and 
with the temperature linearly rising up to $T_1$:
\begin{equation}
  T(x=0)=
  \begin{cases}
    T_0+\frac{T_1-T_0}{\tau}t,&t<\tau\\
    T_1,&t\geq\tau.
  \end{cases}
  \label{eq:left_wall_T}
\end{equation}
The choice of $\tau$ is specified below. $T_1$ is the
temperature which is larger than the temperature of the active burning. All these
quantities depend on the task and are tuned "by hand" after inspection of 
the trial values in each case.
Let $L$ be the size of the region of interest (it is smaller than
the whole region of calculation). The burning leads to the temperature
growth and therefore the pressure rises as well, which in turn generates sound
waves. For the sake of avoiding unnecessary perturbations we should
give time for sound waves to leave the region of interest, i.e. we demand that 
$\tau\gg L/c_s$. In this case the pressure in region $L$ is constant with a good
accuracy. The region $[0; L]$ is the region of interest, that is we
observe flame propagation only there. This region contains uniform
numerical grid and is only a part of the whole calculational domain
$[0; L_0]$, filled with the nonuniform mesh, so that $L\ll
L_0$, $L_0> c_s t_{\rm simul}$, where $t_{\rm simul}$ is the whole
time of simulation. Under these conditions the right boundary does not
play any role in the flame propagation.

Let us describe the numerical method for the solutions of system
(\ref{sys:hydro}). Physically, the system consists
of 3 parts: hydrodynamics, thermo-conductivity, and burning. For  solution
we will use the method of splitting of physical processes. As an example let us 
consider the equation for energy $\epsilon$ in system (\ref{sys:hydro}).
At each time step the equation is split as:
\begin{equation}
  \label{eq:split_scheme}
  \frac{d\epsilon}{dt}=\left(\frac{d\epsilon}{dt}\right)_{\rm
    hydr}+\left(\frac{\partial\epsilon}{\partial t}\right)_{\rm
    nucl}+\left(\frac{\partial\epsilon}{\partial t}\right)_{\rm thermocond},
\end{equation}
\begin{equation}
  \label{eq:split_hydro}
  \left(\frac{d\epsilon}{dt}\right)_{\rm
    hydr}=-\frac{p}{\rho}\frac{\partial v}{\partial x},
\end{equation}
\begin{equation}
  \label{eq:split_nucl}
  \left(\frac{\partial\epsilon}{\partial t}\right)_{\rm
    nucl}=\dot{S},
\end{equation}
\begin{equation}
  \label{eq:split_cond}
  \left(\frac{\partial\epsilon}{\partial t}\right)_{\rm
    cond}=\frac{\partial (\kappa \partial_x T)}{\partial x}.
\end{equation}
These equations are solved in 3 sub-steps:
the hydrodynamic part in which $X_i={\rm const}$ and no energy
generation is assumed; the nuclear part, ($\rho, v={\rm const}$),
and the thermo-conductivity part, ($\rho, X_i, v={\rm const}$).

For the solution of hyperbolic PDE's we use the implicit
lagrangian numerical scheme written in massive coordinates ($ds\equiv\rho
dx$) and described in \citep{SamarskiiPopov}. The quadratic artificial viscosity
is implemented to make the possibility to calculate sharp
discontinuities.
Parabolic PDE (\ref{eq:split_cond}) is solved with Crank--Nicholson numerical scheme.
The kinetic equation for $X_i$ with simplified nuclear network can
be easily integrated analytically for a small time step. The  {\sc aprox13} nuclear network has its own integrator.

Criteria that are used for determination of time and space discretization
of parameters are the following:
\begin{equation}
  \Delta t=\min\limits_i (\Delta t_{\rm courant, i},\Delta t_{\rm nucl, i}),
\end{equation}
where $\Delta t_{\rm courant, i}$ is the Courant condition for $i$-cell and
$\Delta t_{\rm nucl, i}$ is the time step at which all reagent
concentrations change not more than by 1\%.

\subsection{Results of simulations}

Here we  present results of the computations. The approach we use here
is very close to those used in the paper by \cite{TimmesWoosley_ApJ_1992}.
The flame speed is determined by the time evolution of the
position of the front centre, $x(t)$. By definition it is the
point, where $X_{{\rm C}12}=0.5$.
The medium is on the first stage heated with increasing temperature of the wall. At some
moment the temperature rise becomes due to nuclear reactions and the
flame born detaches from the left wall (its subsequent evolution does not
depend on processes at the wall): the temperature of burned
matter exceeds $T_1$, and the characteristic timescale $\tau_{\rm
  nucl}$ of the flame is smaller than $\tau$. For additional study of
independence of results on flame ignition see the Appendix \ref{appa}.

\begin{figure}
 \includegraphics[height=0.50\textwidth,angle=270]{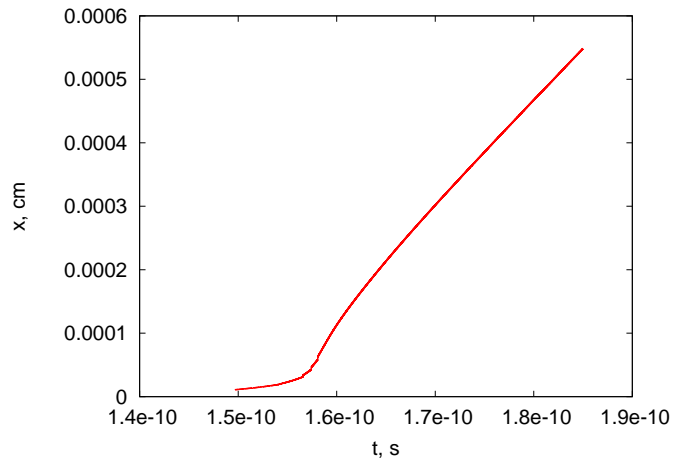}
 \caption{\label{fig:xt} An example of the front coordinate dependence on time 
  for $\rho=2\cdot 10^9$ g/cm$^3$, initial composition ${}^{12}$C, {\sc aprox13} nuclear network is used.}
\end{figure}

\begin{figure}
 \includegraphics[height=0.50\textwidth,angle=270]{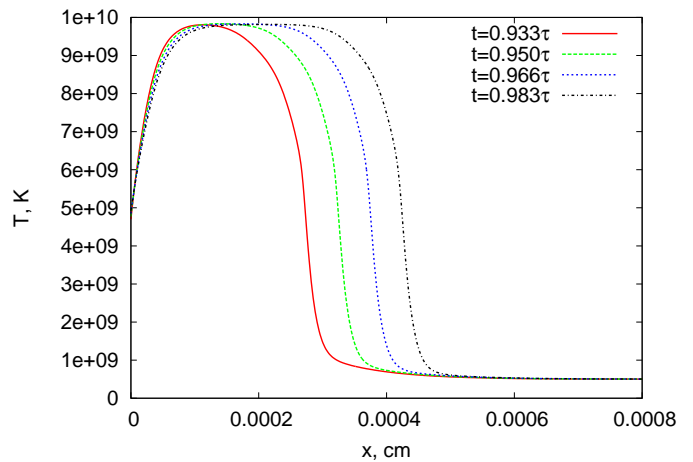}
 \caption{\label{fig:Txts} Sequential profiles of temperature for 
   $t_1<t_2<t_3<t_4$, $\rho=2\cdot 10^9$ g/cm$^3$, initial composition ${}^{12}$C, {\sc aprox13} nuclear network}
\end{figure}

An example of the time dependence, $x(t)$, is
presented in  Fig.~\ref{fig:xt}. In  Fig.~\ref{fig:Txts} the coordinate
dependence of the temperature for a sequence of times is shown.
We see that some time after the ignition the flame
stabilizes and propagates with constant velocity.
Fitting the $x(t)$
dependence we obtain the velocity of flame relative to the
wall, $v$. The normal flame velocity is then $v_n=\rho_b v/\rho_u$. Table \ref{tab:res_C12}
presents the results of the simulations for different initial states of
matter for all nuclear network variants.


\begin{table*}
 \centering
 \begin{minipage}{180mm}
\caption{\label{tab:res_C12} Results of the flame
  simulations. $\rho_0$ -- initial density, ``Calor.'' -- the variant
  of used nuclear network (for details see the text), $T_{\rm max}$ --
  the temperature of burned matter, $\rho_u/\rho_b$ -- the ratio of
  unburned and burned matter densities, $v_n$ -- the normal flame
  velocity, $\Delta x_{\rm fr}$ -- the flame width (determined by characteristic length of temperature rise),
  $v_{\mbox{\tiny TW}}$ --
  flame velocity by Timmes \& Woosley fit functions (should be
  compared with APROX13 velocities). The table also presents
  characteristic timescale of burning $\Delta x_{\rm fr}/v_n$ for our
  results and those by TW ($v_n$ obtained by Eqs. (43), (44) in TW,
  flame width by interpolation of Tables 3--4).}
\begin{tabular}{|c|c|c|c|c|c|c|c|c|c|}
  \hline
  Compos. & $\rho_0$, g/cm$^3$ & Calor. & $T_{\rm max}$, $10^9$ K & $\rho_u/\rho_b$ & $v_n$, km/s & $\Delta x_{\rm fr}$, cm & $\Delta x_{\rm fr}/v_n$, s & $v_{\mbox{\tiny TW}}$, km/s & $(\Delta x_{\rm fr}/v_n)_{\rm TW}$, s \\
  \hline
  $^{12}$C &        & Mg & 6.9 & 1.54 & 70.1 & $1.7\cdot 10^{-4}$ & $2.4\cdot 10^{-11}$ & & \\
        & $2\cdot 10^8$ & Ni & 7.9 & 1.85 & 122 & $1.2\cdot 10^{-4}$ & $9.8\cdot 10^{-12}$ & &\\
        &        & {\sc aprox13} & 6.8 & 1.60 & 18.2 & $3.0\cdot 10^{-4}$ & $1.6\cdot 10^{-10}$ & 26.7 & $5.1\cdot 10^{-10}$ \\
        \cline{2-10}
        &        & Mg & 9.1 & 1.33 & 302 & $1.8\cdot 10^{-5}$ & $6.0\cdot 10^{-13}$ & &\\
         & $7\cdot 10^8$ & Ni & 10.7 & 1.57 & 470 & $1.5\cdot 10^{-5}$ & $3.2\cdot 10^{-13}$ & & \\
         &       & {\sc aprox13} & 8.5 & 1.35 & 55.4 & $4\cdot 10^{-4}$ & $7.2\cdot 10^{-11}$ & 73.2 & $2.9\cdot 10^{-11}$ \\
        \cline{2-10}
          &  & Mg & 11.3 & 1.26 & 854 & $5.5\cdot 10^{-6}$ & $6.4\cdot 10^{-14}$ & & \\
          & $2\cdot 10^9$ & Ni & 13.8 & 1.40 & 1241 & $1.0\cdot 10^{-5}$ & $8.1\cdot 10^{-14}$ & & \\
          &      & {\sc aprox13} & 9.8 & 1.23 & 134 & $1\cdot 10^{-4}$ & $7.5\cdot 10^{-12}$ & 170 & $2.2\cdot 10^{-12}$ \\
  \hline
  $^{16}$O & $2\cdot 10^8$ & {\sc aprox13} & 6.1 & 1.38 & 0.94 & $2\cdot 10^{-2}$ & $2.1\cdot 10^{-7}$ & 2.0 & $1.3\cdot 10^{-7}$ \\
          & $2\cdot 10^9$ & {\sc aprox13} & 8.8 & 1.17 & 22.1 & $5\cdot 10^{-4}$ & $2.3\cdot 10^{-10}$ & 23.0 & $1.5\cdot 10^{-10}$ \\
  \hline
\end{tabular}
\end{minipage}
\end{table*}

\begin{figure}
 \includegraphics[height=0.50\textwidth,angle=270]{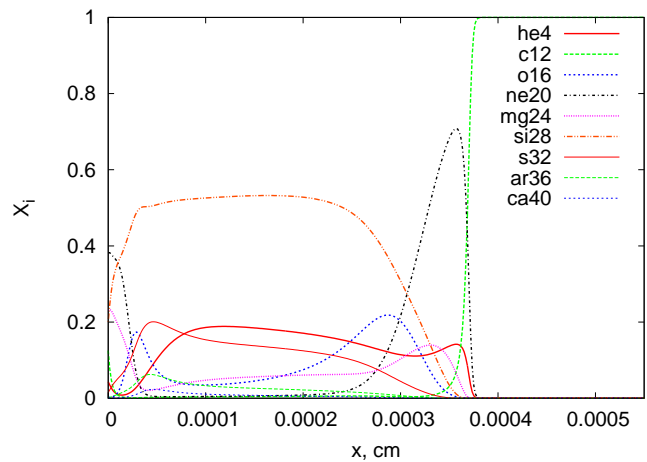}
 \caption{\label{fig:cix} Coordinate dependence of concentrations of
   different elements for {\sc aprox13} nuclear network, $\rho=2\cdot
   10^9$ g/cm$^3$, initial composition ${}^{12}$C}
\end{figure}

Fig.~\ref{fig:cix} shows the
coordinate dependence of $X_i$ for different elements at fixed
time for {\sc aprox13} (flame moves from left to right).
We see that the state close to nuclear quasi-equilibrium is set after
some time of burning (see region $1\cdot 10^{-4}$ to $2\cdot 10^{-4}$
cm on Fig. ~\ref{fig:cix}, the chemical composition near point $x=0$
is perturbed significantly by boundary condition).
The numerically obtained results for 1-step nuclear network are in 
good agreement with analytical estimates (table \ref{tab:theor_vel}). The 
flame speed  obtained with {\sc aprox13} show that the flame speed is about an order of magnitude
slower, compared to one-step reaction.
It is an important result showing that the pure carbon burning to Mg is not a
good approximation, despite the fact that it dominates at early
stages of the burning (see Fig.~\ref{fig:cix}).

The flame speed obtained with {\sc aprox13} (which is supposed to be the real speed) can be calculated analytically if we
substitute  the rate of
reaction $^{24}$Mg$(\alpha,\gamma)^{28}$Si into  Eq.~(\ref{eq:theor_vel})  instead of 2C$\rightarrow$Mg. 
The speed should decrease by
the factor $\sqrt{R_{\rm CMg}(T_9\approx 10)/R_{\rm MgSi}(T_9\approx
  10)}\sim 10$. This result is in accordance with the numerical
simulations. Reactions with $\alpha$ particles dominate the whole burning and have
more or less the same rates (the charge $Z$ of heavy ions in the
governing reactions does not vary significantly), so this estimate is
accurate for all $\alpha$-reactions.

Comparing our
results with those of \cite{TimmesWoosley_ApJ_1992}, here we mean only
comparison with APROX13 nuclear network, we can see
a good (but not excellent) agreement of 20\%--30\% (the worst case is
the lowest density).
Opacities and thermo-conductivities are the same in both
papers. The difference could be caused by different nuclear rates: e.g., the
$^{12}$C$(\alpha,\gamma)^{16}$O reaction rate was changed since TW paper
(F.X.~Timmes, private communication). We have checked the impact of this change on our
results and see no difference: 134 km/s vs 134 km/s for $\rho=2\cdot
10^9$ g/cm$^3$, $^{12}$C run.
Nevertheless, due to uncertainties in Coulomb integral and nuclear
network rates the speed of flame in SNIa is known with accuracy within a few tens of
percent.

\section{Thermal instability of nuclear flames}
\label{sec:th_dyn_ins}

Let us consider now the thermal-pulsational instability with switched-off
hydrodynamics.  Due to absence of sonic waves in this case,
this instability can be described without interference with 
other physical phenomena.
Two conditions are necessary for the development of the TP
instability \citep{Zel,W}. First, the thermal diffusion must be
more than an order of magnitude larger
than the mass diffusion. This condition is undoubtedly satisfied
in presupernovae where ${\rm Le}\gg 1$. Second,
the Zeldovich number \citep{Clav}:
\begin{equation}
\mbox{\sf Ze}=\left(\frac{\partial\ln \dot{S}}{\partial\ln T}\right)_{P,\,
X_i}\ ,
\label{Ze}
\end{equation}
which characterizes 
the temperature dependence of the heating rate $\dot{S}$, 
must be high enough.
Here $P$ is the pressure, $X_i$ ($i=1,2,\dots$) are the abundances of the
reactants, and the derivative is evaluated  at the temperature of the
burned matter (ashes). The thermal instability takes place when {\sf Ze} is
higher than a certain critical value $\mbox{\sf Ze}_{\rm cr}$.
From the Arrhenius law, and in the approximation when the
reaction zone is assumed to be negligibly thin in comparison to the preheat
zone of the flame, it is found that $\mbox{\sf Ze}_{\rm cr}=4+2\sqrt{5}=8.47$ \cite{Zel}.
The recent numerical value \cite{BayMat}, obtained  by  relaxing
the approximation of delta function kinetics,
is $\mbox{\sf Ze}_{\rm cr} \simeq 8.24 - 8.29$. 

\begin{figure}
 \includegraphics[width=0.50\textwidth]{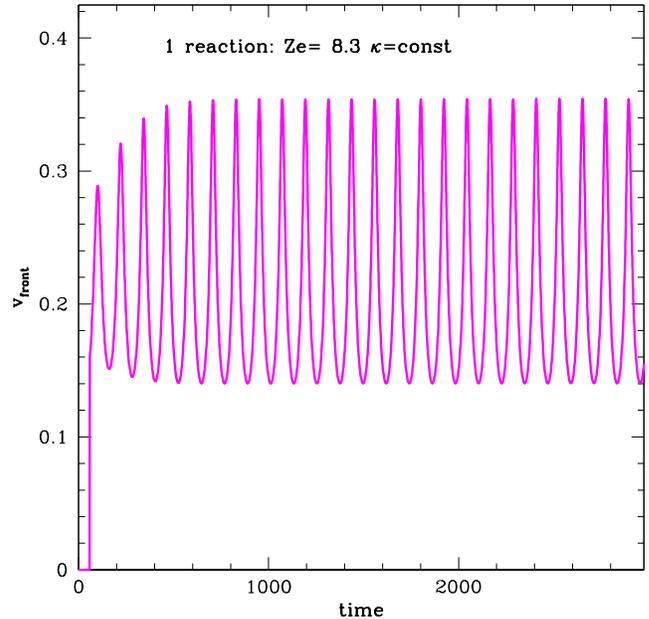}
 \caption{\label{puls} Pulsating front velocity for $\mbox{\sf Ze} = 8.3$ 
and one reaction obeying the Arrhenius law with switched-off hydrodynamics}
\end{figure}

When {\sf Ze} is below the critical
value, the front velocity relaxes quickly to a constant value, but when
it is larger than $\mbox{\sf Ze}_{\rm cr}$ the pulsations set in.
This is illustrated in Fig.\ref{puls}. 

This plot shows the result of the integration of the following
simplified system (compared to Eqns. (\ref{sys:hydro}) $\kappa$ and $B_i$ are taken to be unity):
\begin{equation}
 \frac{\partial T}{\partial t} = \frac{\partial }{\partial x} \left( \frac{\partial T}{ \partial x} \right) + 
  R X_A^2 .
\end{equation}
Here, we assume that the kinetic equation is
\begin{equation}
 \frac{d X_A}{d t} = - R X_A^2 ,
\end{equation}
and the rate of the reaction $R=R(T)$ is taken as
\begin{equation}
 R = \exp\left(-E_a \left(\frac{1}{T}-\frac{1}{T_0+1}\right)\right) .
\end{equation}
In our units the initial temperature is $T_0$, ashes have temperature $T_b=T_0+1$ and
the activation energy is 
\begin{equation}
E_a=\mbox{{\sf Ze}} \frac{T_b}{T_b-T_0} = \mbox{{\sf Ze}} (1+T_0) .
\end{equation}

We have used the method of lines to simulate the solution of the equation of thermal conduction.
Namely, the space coordinate was discretized by finite differences, while
integration over time was done by ODE integrators using methods of
\cite{Gear} and \cite{Brayton}.
See details of the numerical technique in more complicated situation of hydrodynamical
evolution in \cite{BlinnikovDunBark1994}.
Our simulations for this simple model give $\mbox{\sf Ze}_{\rm cr} \simeq 8.2$.


In order to get insight in more complicated simulations of nuclear flames 
discussed in Section~\ref{Discussion}
we have introduced another reactant in our set of equations.

Instead of a large kinetic scheme of many reactions we use a simplified two-reactant system,
which allows us to do simple numerical experiments, elucidating the
role of secondary reactions (``A'' is like carbon, and ``B'' like He - initially He is 0 and appears
as a result of C-burning; the reaction A+B has lower Coulomb
barrier, than A+A, hence lower $E_a$; the system is not closed,
reactions burn to some ashes not included in the system; all
parameters and the system are artificial):
\begin{equation}
 \frac{\partial T}{\partial t} = \frac{\partial }{\partial x} \left( \frac{\partial T}{ \partial x} \right) + 
  \left(\frac{3}{2} -\frac{1}{2} q_2 \right) R_1 X_A^2 + q_2 R_2 X_A X_B .
\end{equation}
and assume the following kinetic equations:
\begin{equation}
 \frac{d X_A}{d t} = - R_1 X_A^2 - R_2 X_A X_B ,
\end{equation}
\begin{equation}
 \frac{d X_B}{d t} = + \frac{1}{2} R_1 X_A^2 - R_2 X_A X_B ,
\end{equation}
where the reaction rates, $R_i=R_i(T)$, are taken as
\begin{equation}
 R_1 = \exp\left(-E_a \left(\frac{1}{T}-\frac{1}{T_0+1}\right)\right) ,
\end{equation}
and
\begin{equation}
 R_2 = \exp\left(- \frac{E_{a2}}{T}+\frac{E_a}{T_0+1}\right) .
\end{equation}
Thus the first reaction has activation energy $E_a$ while the second one has activation energy
$E_{a2}$. 
Changing $E_{a2}$ we are able to study the effect of secondary reaction on the stability 
of the flame, when {\sf Ze}-number is determined by the primary reaction.

Models presented in this section are artificial, but they can show the
variety of effects in systems with burning.

Figs. \ref{fig:ze12}-\ref{fig:ze24} illustrate, that for {\sf Ze}=12 the secondary reaction 
stabilizes the flame, but for large enough {\sf Ze}-number
the instability can develop even when the presence of the secondary reactions.

\FIG{fig:ze12}{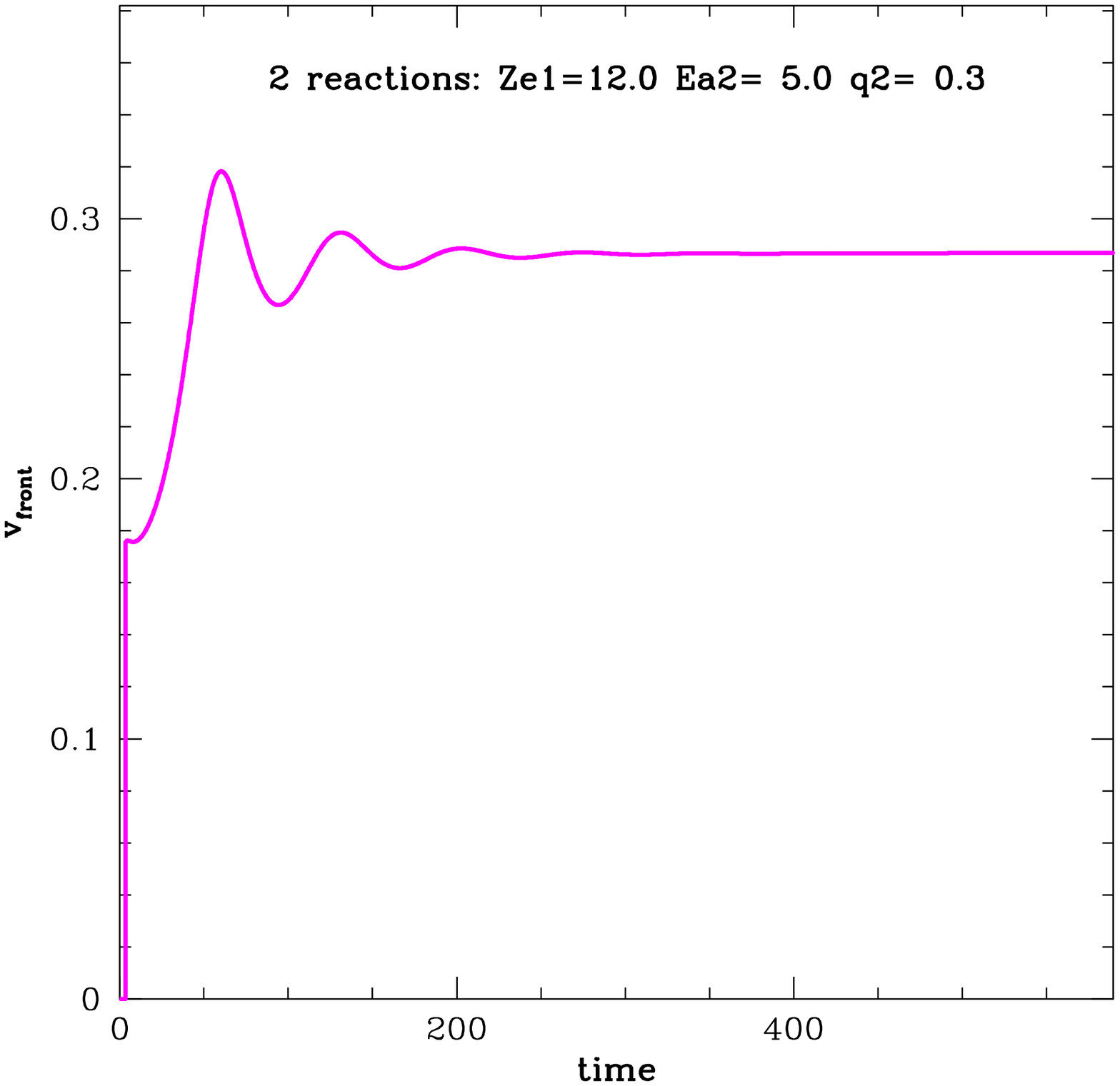}{Pulsating front velocity for $\mbox{\sf Ze} = 12$ 
and a two--reaction artificial network with switched-off hydrodynamics}{width=0.50\textwidth}

\FIG{fig:ze18}{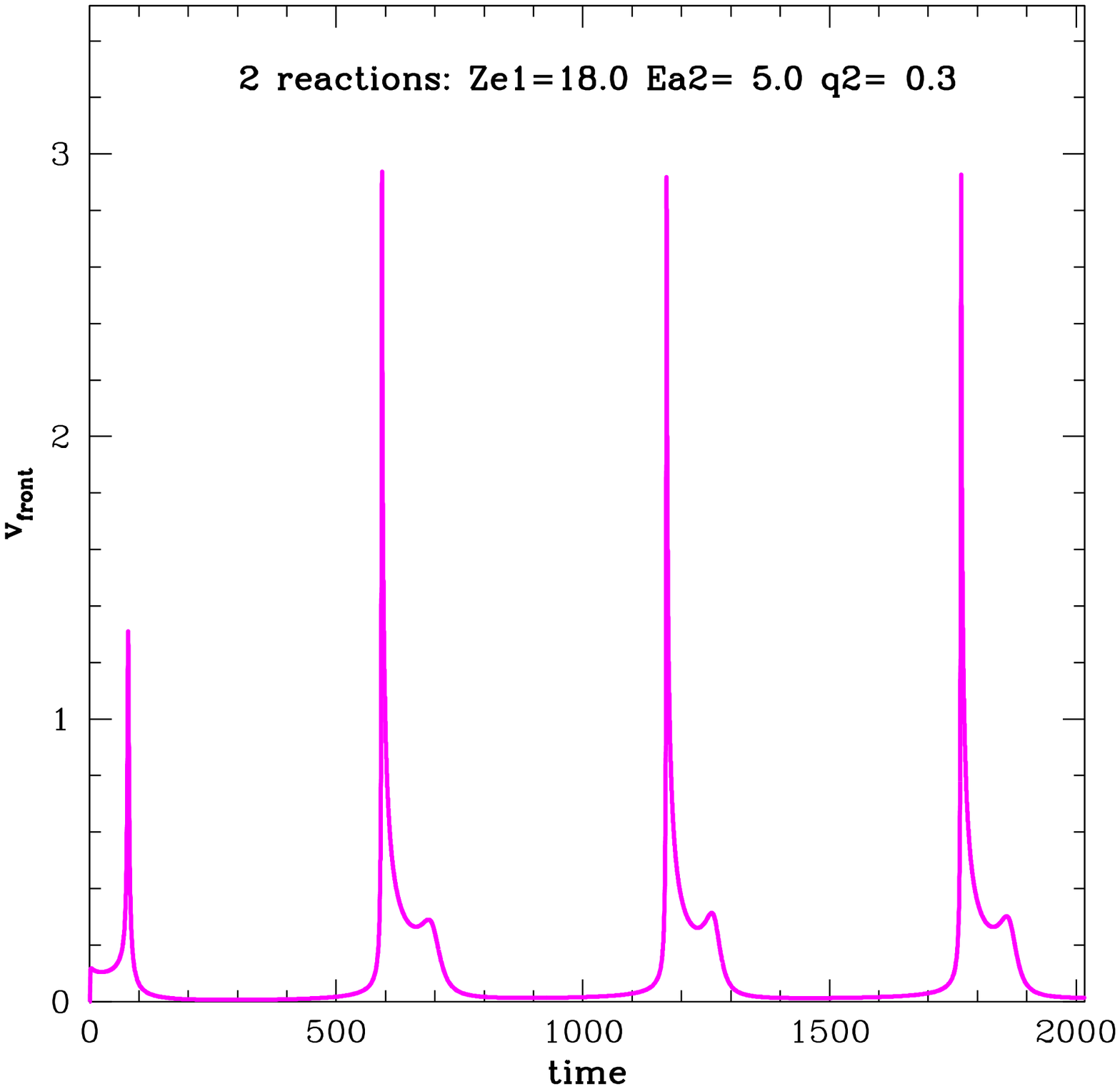}{Pulsating front velocity for $\mbox{\sf Ze} = 18$ 
and a two--reaction artificial network with switched-off hydrodynamics}{width=0.50\textwidth}

\FIG{fig:ze24}{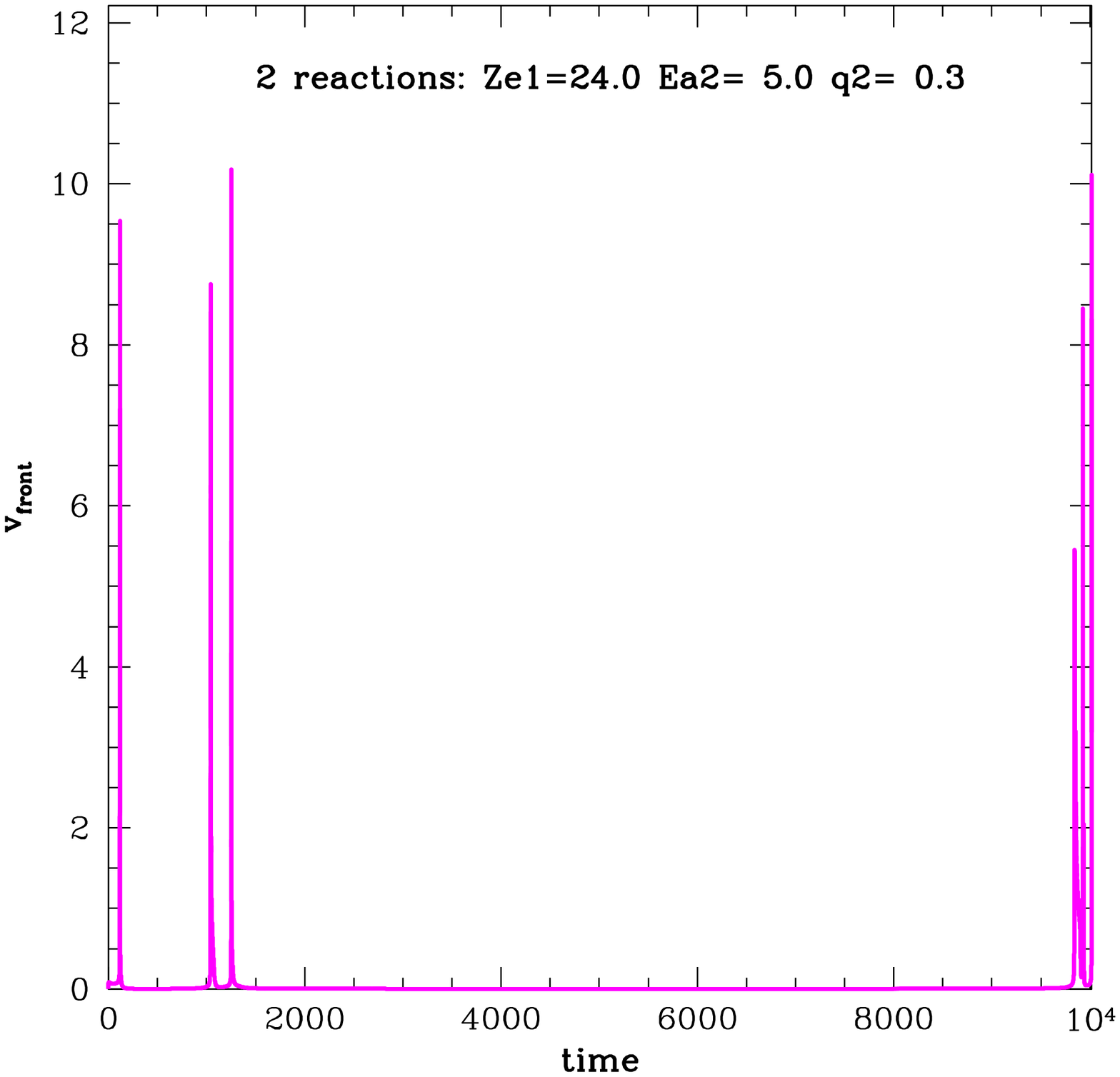}{Pulsating front velocity for $\mbox{\sf Ze} = 24$ 
and a two--reaction artificial network with switched-off hydrodynamics}{width=0.50\textwidth}

Examples in this section were given for the case of the switched-off hydrodynamics.
Now we turn to the flames with an account of hydrodynamics,
which is a better approximation for realistic supernovae. 


\section{Front pulsations}
\label{sec:front_puls}

Here we will switch on hydrodynamics again.
According to the simulations in Section
\ref{sec:flame_prop_1d} with the full set of reactions the instability
 was not observed (see the text below).
Now we show that pulsational behaviour can exist in our simulations
in the case of artificially high Zeldovich numbers and by that will confirm
that carbon and oxygen burning
are stable. Here we should mention that the only source of
fluctuations in our system is numerical noise and we can detect the
pulsational regime only if its frequency is resolved by our numerical
scheme. Looking back to our paper \cite{Sas}, where the results of
numerical simulations of some toy model coincide with analytical
consideration very well, we hope to detect real pulsations in this
system too.
To carry out this task we should change the nuclear rate function.
We treat the reactions as one-step process with the rate given by the Arrhenius law:
\begin{equation}
  R(T)=Ae^{-\B/T_9}.
\end{equation}
Constant $\B$ determines the Zeldovich number, {\sf Ze}, according to eq.
(\ref{Ze}):
\begin{equation}
  {\rm Ze_{Arren}}=\frac{\B}{T_{9~\rm burned}}\,,
\end{equation}
where $T_{9~\rm burned}$ is
the temperature of the burned matter (ashes).

One of the goal of this Section is to show the possibility to resolve
pulsations in numerical codes like ours. The proposed simplified
one--step reaction with Arrhenius rate is the way to parameterize
the Zeldovich number. Changing $\B$ we simply change {\sf Ze}.

The value of constant $A$ does not play any role here, but for
definiteness we determine it by the relation:
\begin{equation}
  \int\limits_{T_1}^{T_2}R(T_9)dT_9=\int\limits_{T_1}^{T_2}R_{\rm real}(T_9)
  dT_9,
\end{equation}
\begin{equation}
  A(\B)=\int\limits_{T_1}^{T_2}R_{\rm real}(T_9)dT_9\left(\int\limits_{T_1}^{T_2}e^{-\B/T_9}dT_9\right)^{-1}.
\end{equation}
For $R_{\rm real}(T)$ we take the rate of the considered above 
simplified nuclear network.

The nuclear caloricity is taken from nuclear transformations
$^{12}$C$\rightarrow^{24}$Mg and $^{12}$C$\rightarrow^{56}$Ni
(the used value will be denoted in each simulation explicitly in the
data table).
According to the results of the previous
section we set the boundary temperatures as $T_1=1$, $T_2=10$.
Fig. \ref{fig:rates_cmp} shows different rates used, compared  to the ``real
rate''. We see that the rate with $\B\approx 50$ fits the physical carbon burning
rate very nicely. This means that when $T_{9 {\rm up}}\approx 10$,
${\rm {\sf Ze}}\approx 5$, which coincides with estimates presented in
Section \ref{Discussion}.
Table \ref{tab:Arr_runs} presents the results for different runs with
$\rho_0=2\cdot 10^9$ g/cm$^3$. We increase $\B$
from 20 to 200 and simultaneously decrease  caloricity $q$ (from Ni to
Mg) to avoid the flame acceleration. We see that for the normal caloricity
($q=(5.6-9.2)\cdot 10^{17}$~erg/g) no pulsations appear for all
variants of rates (upper part of the table).
Zeldovich number, ${\rm {\sf Ze}}$, can be additionally increased by decreasing $T_{9\rm burned}$. This is done by
the caloricity change to $q=5-7$ MeV for one reaction ($q=(2.0-2.8)\cdot
10^{17}$ erg/g). In this case
pulsations appear for $\B>\B_{\rm crit}\simeq 112.5$ (the lower part of the table). Example of the flame
coordinate dependence is depicted in Fig.~\ref{fig:xt_puls}, and the
corresponding evolution of temperature in
Fig.~\ref{fig:T_puls}. Additional variants were calculated for
the intermediate caloricity $q=3.5\cdot 10^{17}$ erg/g. The results for
this caloricity show the same Zeldovich number (within the
uncertainties of results).

\FIG{fig:rates_cmp}{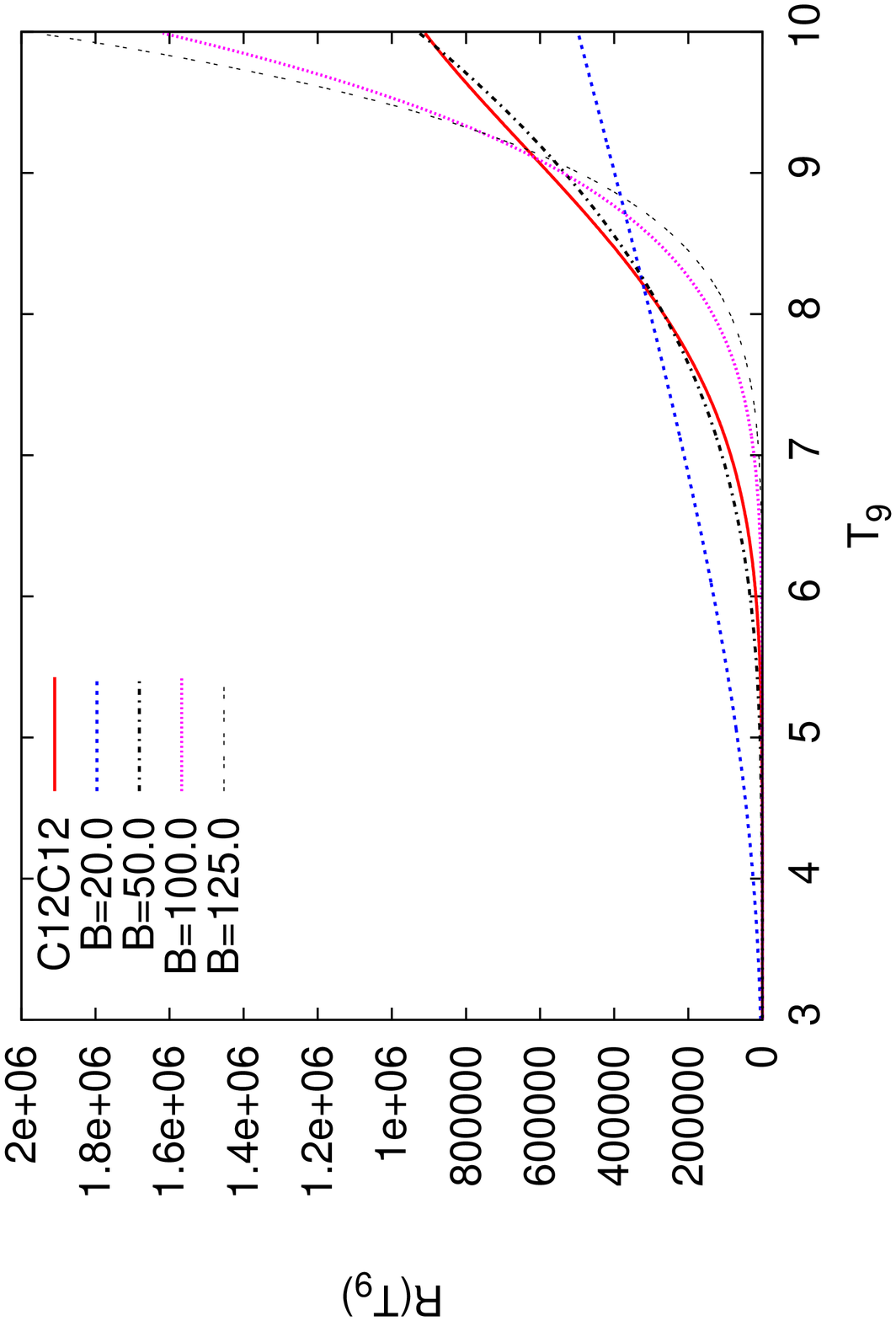}{Comparison of artificial rates with
  2$^{12}$C$\rightarrow$Mg$^{24}$ (in the plot keys B is $\B$)}{height=0.50\textwidth,angle=270}

\FIG{fig:xt_puls}{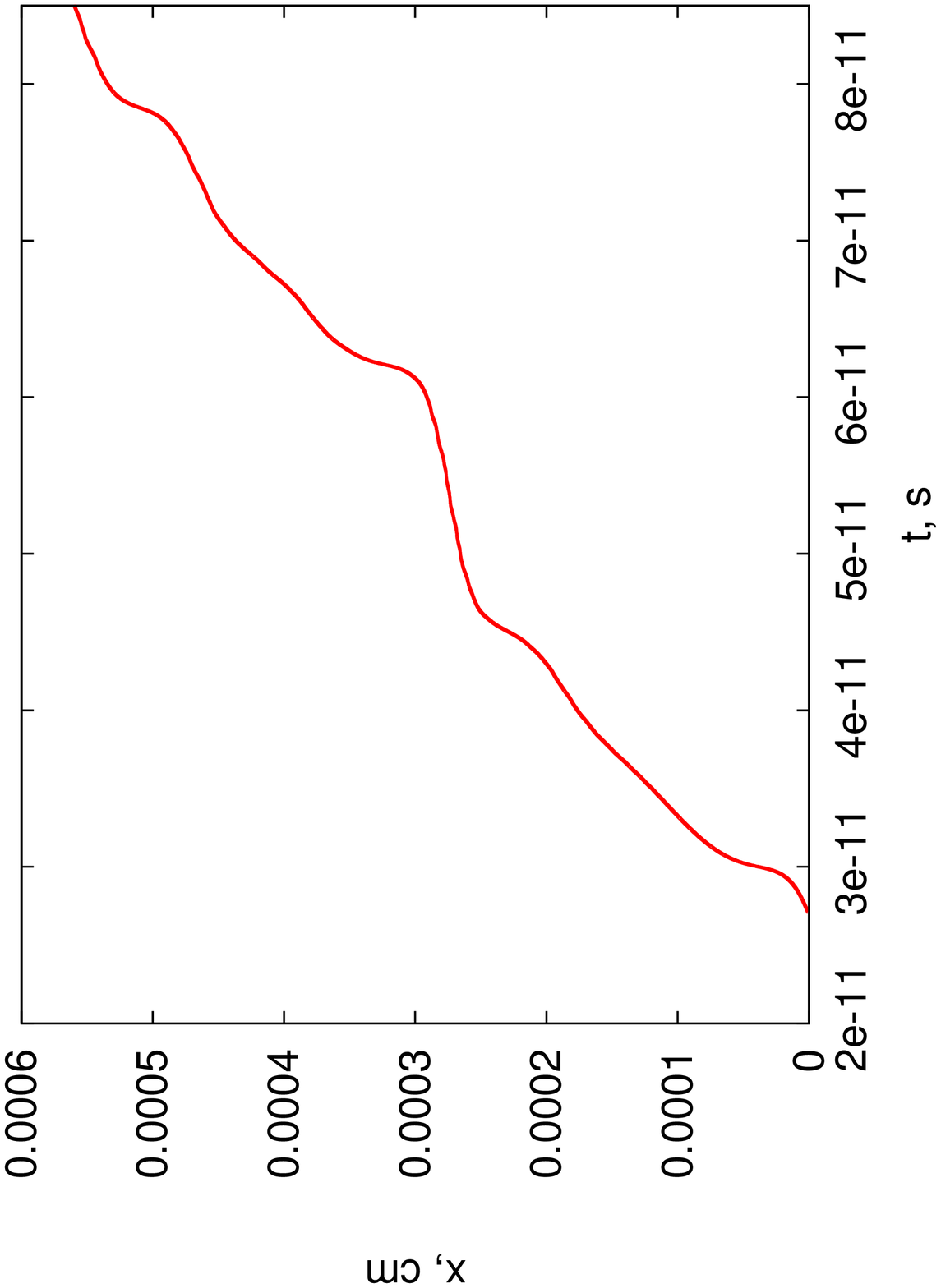}{Pulsational regime of flame with
  artificially increased Zeldovich number, $\rho=2\cdot 10^9$
  g/cm$^3$, $\B=112.5$, $q=2.4\cdot 10^{17}$ erg/g}{height=0.50\textwidth,angle=270}

\FIG{fig:T_puls}{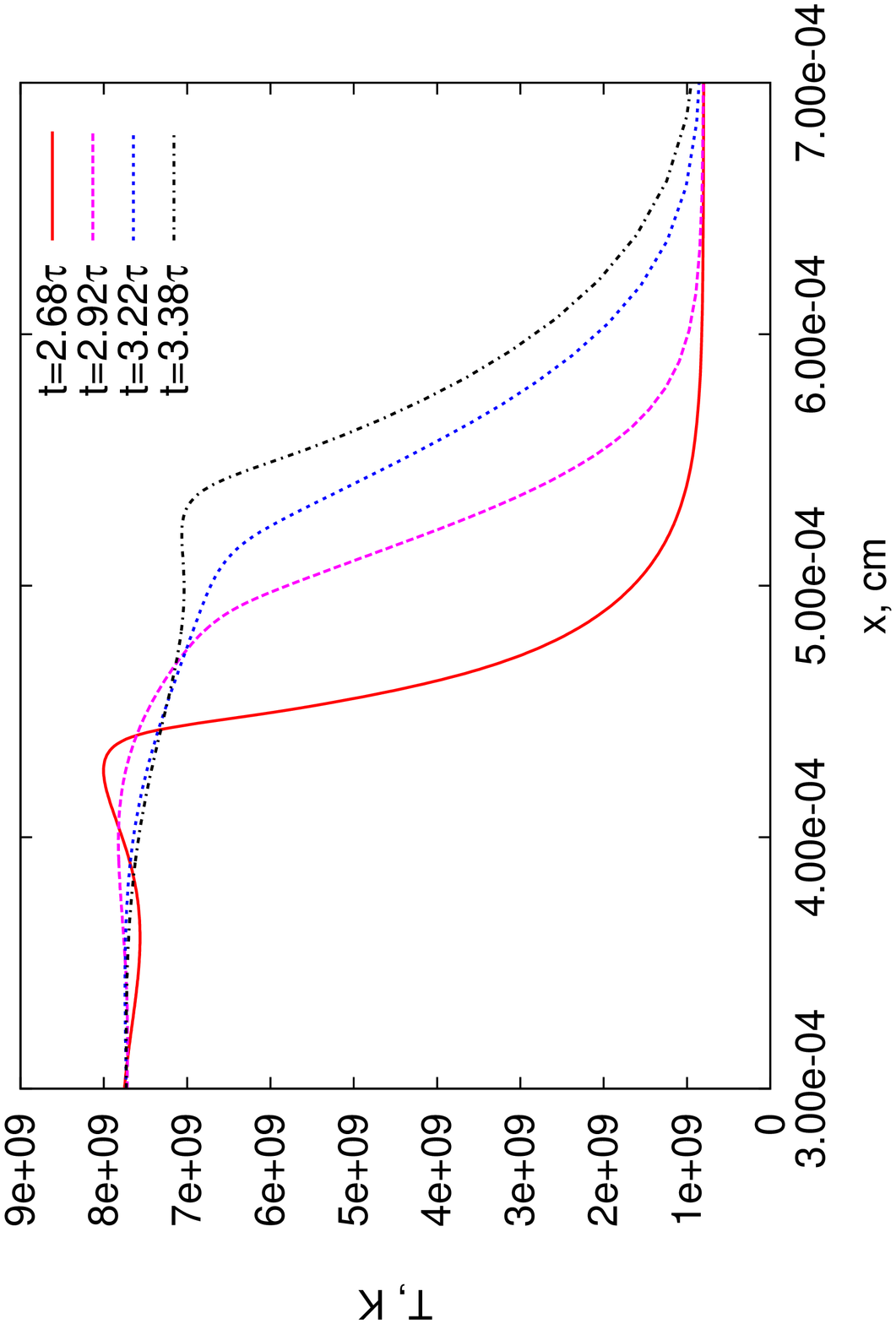}{Temperature profiles for pulsational regime of flame with
  artificially increased Zeldovich number, $\rho=2\cdot 10^9$
  g/cm$^3$, $\B=112.5$, $q=2.4\cdot 10^{17}$ erg/g}{height=0.50\textwidth,angle=270}

So the result is that in ``real''  system (physical EOS and
thermoconductivity) pulsations can exist, but only in the
case of unphysical values of the parameters. In particular,
the Zeldovich number should be
increased by a factor of $\sim 4$. Table \ref{tab:Ze_rho} presents
the critical values of {\sf Ze}  number found in our numerical experiments for different densities of
matter.

In the case of complex nuclear network (many reactions) {\sf Ze} 
number cannot be obtained in such straightforward way, because it depends on the
whole history of burning $\dot{S}(T)$. The front is stable in simulations with {\sc aprox13},
presented in Section \ref{sec:flame_prop_1d} for all runs.
The only run which indicates the instability,
but in reality is stable, is with initial $^{16}$O and
$\rho=2\cdot 10^8$ g/cm$^3$ (one from the bottom in Table~\ref{tab:res_C12}). It demonstrates
some irregular structure in $x(t)$
dependence (see Fig. \ref{fig:xt_puls2} and compare to  Fig. \ref{fig:xt_puls}).
This case is the best candidate for
pulsations because of low $T_b$ and strong temperature dependence. But
the period of these pulsations, $\tau_1\sim 3\cdot 10^{-10}$ s, is not
far from the numerical cell crossing time $\tau_2=\Delta x
\rho_b/(u_n\rho_u)=1.5\cdot 10^{-9}$~s (here $\Delta x$ is the size of
a cell), and to check, whether it is a
numerical effect, we perform simulation with different $\Delta x$, see
Fig.~\ref{fig:xt_puls2}. It could be seen that the period of
pulsations decreases with the cell size together with the mean squared
deviation from the linear law. So these pulsations are numerical
and they do not contribute to physical effects like changing
the flame velocity.
Another signature of physical pulsations (not observed in this run),
is that they appear not only in $x(t)$ dependence and also in
pulsations of $T_{\rm burned}$ in the same way as was shown in the
paper by \cite{Sas}.

\FIG{fig:xt_puls2}{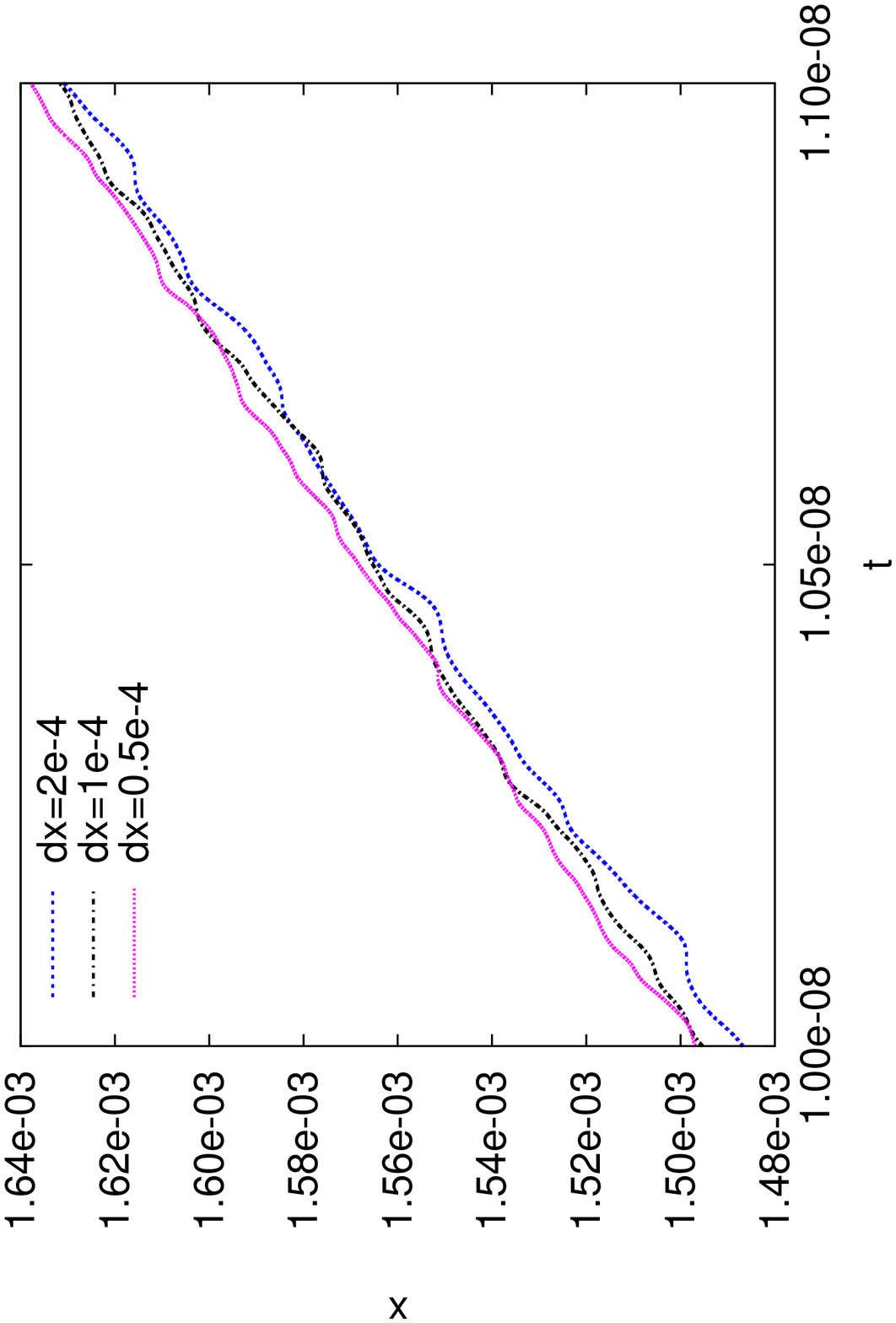}{Flame coordinate for run with
  $^{16}$O, $\rho=2\cdot 10^8$ g/cm$^3$, {\sc aprox13} network. Three
  runs with the same conditions but different resolution (cells size).}{height=0.48\textwidth,angle=270}

\begin{table}
\caption{\label{tab:Arr_runs}Results of simulations with Arrhenius law:}
\begin{center}
\begin{tabular}{|cccccc|}
  \hline
  № & $A$ & $\B$ & $q$, $10^{17}$ erg/g & $T_{9\rm burned}$ & comm. \\
  \hline
  1 & $3.68\cdot 10^6$ & $20.0$ & 9.2 & $13.5$ & flame \\
  2 & $1.38\cdot 10^8$ & $50.0$ & 9.2 & $13.5$ & flame \\
  3 & $7.62\cdot 10^{12}$ & $150.0$ & 5.6 & $11$ & flame \\
  4 & $1.47\cdot 10^{15}$ & $200.0$ & 5.6 & $14$ & deton. \\
  \hline
  5 & $7.62\cdot 10^{12}$ & $150.0$ & 2.8 &  $10$  & puls. \\
  6 & $3.60\cdot 10^{10}$ & $100.0$ & 2.8 & $8$ & flame \\
  7 & $5.32\cdot 10^{11}$ & $125.0$ & 2.8 & $8$ & puls. \\
  8 & $1.39\cdot 10^{11}$ & $112.5$ & 2.8 & $8.2$ & flame \\
  9 & $1.39\cdot 10^{11}$ & $112.5$ & 2.0 & $7.8$ & puls. \\
  10 & $1.39\cdot 10^{11}$ & $112.5$ & 2.4 & $8$ & puls. \\
  11 & $4.44\cdot 10^{11}$ & $123.3$ & 3.5 & $9.0$ & puls.\\
  12 & $3.11\cdot 10^{11}$ & $120.0$ & 3.5 & $9.0$ & flame\\
  \hline
\end{tabular}
\end{center}
\end{table}

\begin{table}
\caption{\label{tab:Ze_rho}Critical Zeldovich number for carbon
  burning at different densities (parameterized with the one-step reaction rate):}
\begin{center}
\begin{tabular}{|cc|}
  \hline
  $\rho$, g/cm$^3$ & ${\rm Ze_{cr}}$ \\
  \hline
  $2\cdot 10^8$ & $18.4<{\rm {\sf Ze}}<20.4$ \\
  $7\cdot 10^8$ & $14.6<{\rm {\sf Ze}}<15.6$ \\
  $2\cdot 10^9$ & $13.5<{\rm {\sf Ze}}<13.9$ \\
  \hline
\end{tabular}
\end{center}
\end{table}

\section{Discussion}
\label{Discussion}

\citet{TimmesWoosley_ApJ_1992} simulated  the flame fronts of
different CO mixtures using one-dimensional time-dependent
hydro code. They said nothing about the thermal-pulsational (TP) instability. At first
glance this means that there is no TP instability in the
degenerate presupernova conditions. However, later,
\citet{BL}  and 
\citet{Nomoto95} have argued
that C+C and O+O flame fronts should be TP-unstable.
In the approximation of delta function kinetics it is found in \cite{BL}
that $\mbox{\sf Ze}_{\rm cr}=8+4\sqrt{5}=16.9$. It is 
larger than $\mbox{\sf Ze}$ calculated by the Arrhenius law  exactly
by factor  2 (and close to ours in Table \ref{tab:Ze_rho}). 
\cite{BL} even claimed 
that 
\citet{TimmesWoosley_ApJ_1992} have overlooked the instability
in their simulations.

We believe that the latter extreme proposition is not true and
based on the results of the simulations presented above we put forward 
a different explanation of the apparent CO front stability.
It is known that CO nuclear burning consists of the basic (C,$\alpha$),
(C,p), (O,$\alpha$), ... reactions accompanied by secondary (p,$\gamma$)
and ($\alpha$,$\gamma$) reactions. It is very important that:
\begin{enumerate}
\item
 {\sf Ze} corresponding to the secondary reactions is rather
low because of the lower height of the Coulomb barrier;
\item
The amount of energy yielded by the secondary reactions
is comparable to the yield of the basic ones.
\end{enumerate}
It is very difficult to study analytically any model with many
reactants. It should be made numerically as it was done in  Section
\ref{sec:flame_prop_1d} and by \citet{TimmesWoosley_ApJ_1992}. 
However, it is possible to develop a simple analytical
model like in \cite{Sas}, \cite{Weber} with {\em one} deficient reactant which possesses in part the
main properties of CO burning mentioned above, and which can help 
to understand the qualitative behaviour of the pulsating front. 
Moreover, as we have shown in Section \ref{sec:th_dyn_ins},
it is not hard to develop a numerical model with only {\em two} reactions
with different values of activation energies giving two {\sf Ze} numbers, 
one of them above and the other
below the critical value. This model 
explains how the  seeming contradiction between works~\cite {BL,Nomoto95}, 
and the simulations in Ref.~\cite{TimmesWoosley_ApJ_1992} can be removed.

The results of Section \ref{sec:front_puls} show that for physical
parameters of a white dwarf matter it is possible to obtain a
pulsational regime of the flame propagation, but only in case of
increased Zeldovich number. The Zeldovich number of a one-step
reaction with Arrhenius law reflects the dynamics with similar net
Zeldovich number (obtained by real $\dot{S}(T)$ history over flame) of
a more complex network like aprox13.
We determined the {\sf Ze} number in results 
obtained by \cite{TimmesWoosley_ApJ_1992} (see their Fig. 3), evaluating the ratio of the thickness
of the conductive zone to the thickness of the reaction zone which reflects
the effective value of the {\sf Ze} number \citep{Clav}.
We find this value in the range 2 -- 5. This range is noticeably lowerer the critical
value ${\mbox{\sf Ze}}_{\rm cr}$.
This fact suggests the explanation for the 
stability of CO flame front observed by \citet{TimmesWoosley_ApJ_1992}.
Moreover, it shows that the secondary reactions of CO burning are very
important from the viewpoint of the TP stability
and that the models considered in
\citep{Nomoto95,BL} are  oversimplified.

\cite{TimmesWoosley_ApJ_1992} suggested electron captures
as a stabilizing effect for 
the Rayleigh--Taylor instability. This effect do not have impact on
pulsational instability because electron capture is a process with the
time scale of weak processes $>10^{-4}$ s. And the timescale of the
pulsational instability is much smaller (see Table~\ref{tab:res_C12}).

\section{Conclusions}

One-dimensional flame propagation in presupernova
white dwarf has been considered. Flame properties for different star densities were
obtained. It is shown that when only one nuclear reaction
in the nuclear network is considered, the flame velocity strongly differs
from more sophisticated net simulations. So this simplified approach
can be used only in approximate simulations when the exact value of the flame
velocity is not required, or alternatively the nuclear rate should be
adjusted to fit the correct value of $v_n$.

We have also studied one-dimensional pulsational instability.
First, with artificial systems the switched--off hydrodynamics were
considered. The possibility of secondary reactions to stabilize
the front was shown with the help of this system (by decreasing the net {\sf Ze} number).

In ``real'' simulations, presented in Section \ref{sec:flame_prop_1d} no pulsation regime was found,
so we used the simplified
Arrhenius law in one-step reaction to make it possible to change
Zeldovich number, {\sf Ze}. For high {\sf Ze} numbers the pulsations
do exist. It means that our numerical code can resolve such
pulsations and that they can exist in conditions close to ``real'',
but with steeper energy generation rate.
By means of numerical simulations we have obtained critical Zeldovich numbers for densities in the range
$\rho=[2\cdot 10^8, 2\cdot 10^9]$ g/cm$^3$. These values are larger
than those in real flame, which proves one-dimensional stability of the realistic flame fronts in supernovae.

\bigskip
\centerline{\bf Acknowledgements}
We are grateful to V.Chechetkin, W.Hillebrandt, A.Kruzhilin, J.Niemeyer,
P.Sasorov, S.Woosley, and F.Timmes for cooperation, discussions and references.
MPA supported the work of SB in Germany.
The work in Russia is supported by RFBR grants
11-02-00441-a and 13-02-92119, Sci. Schools 5440.2012.2, 3205.2012.2 and 3172.2012.2,
 by the contract No.~11.G34.31.0047 of the Ministry of Education and
 Science of the Russian Federation, and SCOPES project No.~IZ73Z0-128180/1.

\bibliographystyle{apsrev}
\bibliography{front_params}

\appendix
\section{Parameters of simulations}
\label{appa}

\begin{table}
\caption{\label{tab:params_sim}Parameters of simulations:}
\begin{center}
\begin{tabular}{|ccccc|}
  \hline
  $\rho$, g/cm$^3$ & $T_0$, K & $T_1$, K & $L$, cm & $\tau$, s \\
  \hline
  $2\cdot 10^8$ &               &              & $9\cdot 10^{-3}$ &                  \\
  $7\cdot 10^8$ & $5\cdot 10^8$ & $5\cdot 10^9$ & $2.4\cdot 10^{-3}$ & $1.8\cdot 10^{-10}$ \\
  $2\cdot 10^9$ &               &              & $4.5\cdot 10^{-4}$ &                    \\
  \hline
\end{tabular}
\end{center}
\end{table}
The parameters of all simulations are presented in
Table~\ref{tab:params_sim}. The characteristic time of wall heating
$\tau$ (see Eq. \ref{eq:left_wall_T}) has no impact on the results
because the flame is born at times $t>\tau$ (except the variant when
$\rho=2\cdot 10^9$ g/cm$^3$, where $t\sim\tau$, but for this variant
the flame transition time satisfies $L/v_n\ll\tau$, so it crosses the
region of interest $L$ with constant $T_{\rm left}$).
We have made several additional runs to check the dependence of our
results on the initial and boundary temperatures. The results are presented
in Table~\ref{tab:add_runs} and show weak dependence of velocity on
the initial temperature. It happens because higher $T_0$ leads to
faster heating of unburned matter by thermo-conductivity to the
temperature of active burning and by this leads to small increase of
flame velocity.
\begin{table}
\caption{\label{tab:add_runs}Results of additional runs for testing
  impact of run parameters. The second column shows the parameter that
  was changed together with its new value. $(v_n)_{\rm new}$ and
  $(v_n)_{\rm old}$ are new and old values of normal flame velocity,
  respectively, ${}^{12}$C initial composition, aprox13 network.}
\begin{center}
\begin{tabular}{|cccc|}
  \hline
  $\rho$, g/cm$^3$ & param. change & $(v_n)_{\rm new}$, km/s & $(v_n)_{\rm old}$, km/s \\
  \hline
  $2\cdot 10^8$ & $T_0\rightarrow 7\cdot 10^8$~K & 18.6 & 18.2 \\
  & $T_1\rightarrow 4\cdot 10^9$~K & 18.2 & 18.2 \\
  $2\cdot 10^9$ & $T_0\rightarrow 7\cdot 10^9$~K & 136 & 134 \\
  & $T_1\rightarrow 4\cdot 10^9$~K & 134 & 134 \\
  \hline
\end{tabular}
\end{center}
\end{table}

\end{document}